\documentclass[journal,twoside,web]{ieeecolor}
\usepackage{tmi}
\makeatletter
\let\NAT@parse\undefined
\makeatother
\usepackage[colorlinks]{hyperref}
\usepackage{cite}
\usepackage{multirow}
\usepackage{amsmath,amssymb,amsfonts}
\usepackage{algorithmic}
\usepackage{graphicx}
\usepackage{textcomp}
\usepackage{lineno}

\def\BibTeX{{\rm B\kern-.05em{\sc i\kern-.025em b}\kern-.08em
    T\kern-.1667em\lower.7ex\hbox{E}\kern-.125emX}}
\begin{document}
%\linenumbers
\title{M2MRF: Many-to-Many Reassembly of Features for Tiny Lesion Segmentation in Fundus Images}
\author{Qing Liu, Haotian Liu, Wei Ke, Yixiong Liang
\thanks{This work is supported in part by the High Performance Computing Center of Central South University. Qing Liu is partially supported by the National Natural Science Foundation of China (NSFC No. 62006249), Changsha Municipal Natural Science Foundation (kq2014135) and Hunan Provincial Natural Science Foundation of China (2021JJ40788). Wei Ke is partially supported by the National Natural Science Foundation of China (NSFC No. 62006182). Yixiong Liang is partially supported by the Hunan Provincial Science and Technology Innovation Leading Plan (No.2020GK2019). }
\thanks{Qing Liu, Haotian Liu and Yixiong Liang are with School of Computer Science, Central South University, Changsha, Hunan, 410083, China (email: qing.liu.411@gmail.com, (yxliang, haotianliu)@csu.edu.cn). Wei Ke is with Xi'an Jiaotong University,  Xi'an, Shaanxi, China (email: wei.ke@xjtu.edu.cn). Qing Liu and Haotian Liu are equal contribution. Wei Ke and Yixiong Liang are co-corresponding authors.}}

\maketitle

\begin{abstract}
	Feature reassembly is an essential component in modern CNN-based segmentation approaches, which includes feature downsampling and upsampling operators.  Existing operators reassemble multiple features from a small predefined region into one for each target location independently. This may result in loss of spatial information, which could vanish activations caused by tiny lesions particularly when they cluster together. In this paper, we propose a many-to-many reassembly of features (M2MRF). It reassembles features in a dimension-reduced feature space and simultaneously aggregates multiple features inside a large predefined region into multiple target features. In this way, long range spatial dependencies are captured to maintain activations on tiny lesions. Experimental results on two lesion segmentation benchmarks, i.e. DDR and IDRiD, show that (1) our M2MRF outperforms existing feature reassembly operators; (2) equipped with our M2MRF, the HRNetv2 is able to achieve significant better performance to CNN-based segmentation methods and competitive even better performance to two recent transformer-based segmentation methods. Our code is made publicly available at \url{https://github.com/CVIU-CSU/M2MRF-Lesion-Segmentation}.
\end{abstract}

\begin{IEEEkeywords}
	feature reassembly, upsampling operator, downsampling operator, tiny lesion segmentation, fundus image analysis
\end{IEEEkeywords}

% \fi
\IEEEpeerreviewmaketitle

\section{Introduction}
\label{sec:introduction}
\begin{figure}
	\begin{center}		
		\includegraphics[width=0.5\textwidth]{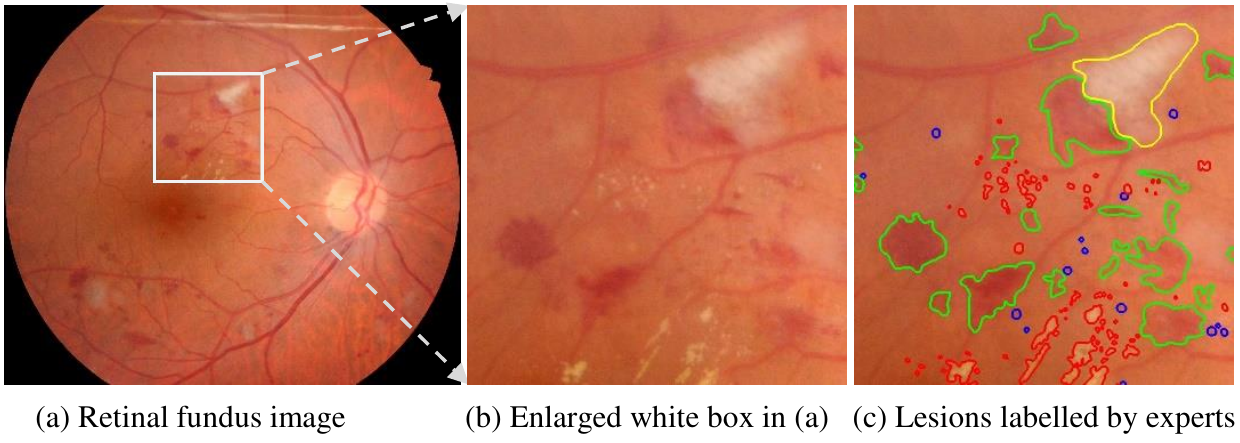}	
	\end{center}
	\caption{Example for colour fundus images and lesions. Four types of
		lesions i.e. SEs, EXs, MAs and HEs are delineated in yellow, red, blue
		and green respectively. Many lesions are extremely small and the size
		variance across them are enormous}
	\label{fig:taskIntroduce}
\end{figure}

\IEEEPARstart{T}{his} paper focuses on segmentation of tiny lesions, e.g. soft exudates (SEs), hard exudates (EXs), microaneurysms (MAs) and hemorrhages (HEs) (see Fig. \ref{fig:taskIntroduce}) in colour fundus images, which is an important prerequisite for enabling computers to assist human doctors for clinical purpose. It falls into the research area of semantic segmentation.

With the rise of deep convolutional neural networks (CNN), many CNN-based approaches such as \cite{long2015fully, ronneberger2015u, xie2015holistically, chen2017deeplab, zhao2017pyramid, chen2018encoder, wang2020deep} have been proposed for semantic segmentation and achieved extraordinary progress. On Cityscapes \cite{cordts2016cityscapes} with 19 classes, the mean of class-wise intersection over union (mIoU) by a state-of-the-art approach HRNetV2 \cite{wang2020deep} has been improved to 81\%. However, when fine-tuning HRNetV2 \cite{wang2020deep} for four-class lesion segmentation in fundus images, the mIoU on IDRiD \cite{porwal2020idrid} decreases to 47\%. Why is lesion segmentation so much harder than natural scene image segmentation?

Two possible factors behind performance gap are the extreme small size of lesions and large scale variation across them. Interestingly, almost 50\% lesions are less than 256 pixels in images of size $4288\times 2848$ from IDRiD \cite{porwal2020idrid} dataset provided by the 2018 ISBI grand challenge. Such tiny size of lesions rises an extreme challenge for CNN-based segmentation approaches to learn discriminative representations with enough spatial information. To make matters worse, the smallest 10\% lesions in IDRiD \cite{porwal2020idrid} only contain less than 74 pixels while the largest 10\% lesions contain more than 1920 pixels. Tiny lesions require CNNs maintain as enough local information as possible within a small receptive field while large lesions require CNNs exploit long range contexts over a large receptive field. This demand contradiction raised by enormous size variation presents another challenge for CNN-based approaches. Moreover, in fundus images, multiple lesions with large size variation always coexist and cluster together, which makes the representation learning more challenge.

\begin{figure*}
	\begin{center}		
		\includegraphics[width=0.85\textwidth]{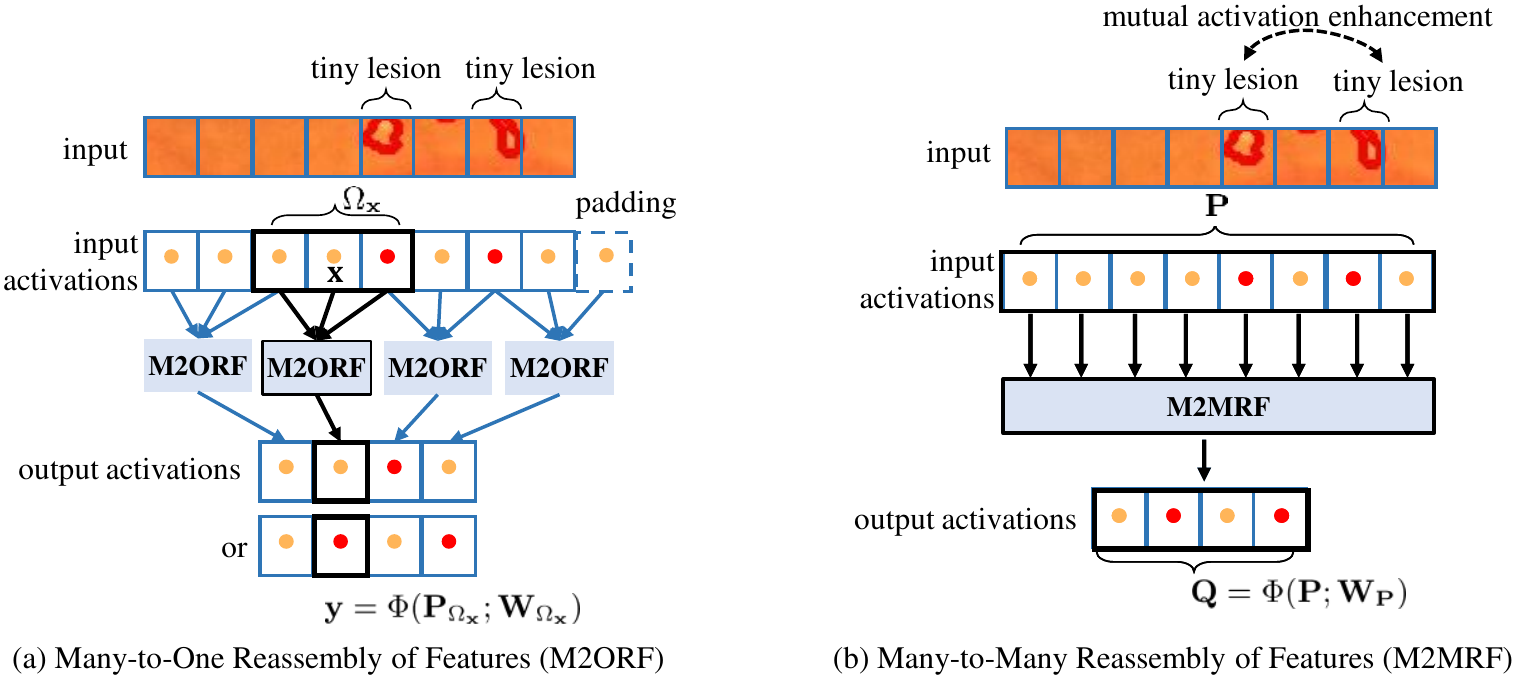}	
	\end{center}
	\caption{Illustration of our motivation. Here we take the downsampling with rate of 2 for example. Input contains two EXs, which are delineated with red lines. Feature activations mainly caused by background and EXs are represented by circles in light orange and red respectively.  With the input activations, we want to preserve activations of both EXs. To generate activation $\mathbf{y}$, M2ORF first finds the source feature $\mathbf{x}$ in input activations, and then determines a small local region $\Omega_{\mathbf{x}}$ centred on $\mathbf{x}$, and finally reassembles features inside $\Omega_{\mathbf{x}}$ with importance weighting kernel $\mathbf{W}_{\Omega_{\mathbf{x}}}$ into target activation $\mathbf{y}$, as illustrated in (a). M2ORF ignores the coexist of multiple lesions, which may vanish tiny lesion activations. Instead, our M2MRF directly learns to reassemble activations inside a large local region $\mathbf{P}$ to generate multiple output activations $\mathbf{Q}$ simultaneously which enables subtle activations be enhanced mutually and well preserved, as illustrated in (b). Here $\mathbf{W}_{\mathbf{P}}$ is the importance weighting kernel in M2MRF.}
	\label{fig:M2M-M2One}
\end{figure*}

Taking an in-depth look at modern CNN-based segmentation approaches \cite{long2015fully, ronneberger2015u, xie2015holistically, chen2017deeplab, zhao2017pyramid, chen2018encoder, sun2019deep, wang2020deep}, the challenges of applying them to lesion segmentation are mainly caused by the repetitive utilisation of operators for reassembly of features (RF operators for short), i.e. downsampling and upsampling operators. Downsampling operator is the key contributor to enlarge CNNs' receptive fields via reducing the spatial resolution of feature maps \cite{chen2017deeplab, gao2019lip}. Conversely, upsampling operator is essential to recover the spatial resolution for better segmentation. Strided max-pooling and strided convolution \cite{springenberg2014striving, sun2019deep, wang2020deep} are two widely-adopted operators for downsampling while bilinear interpolation and deconvolution \cite{zeiler2010deconvolutional}\cite{ zeiler2014visualizing, long2015fully, noh2015learning} are two common choices for feature upsampling. Those RF operators assume that features in target feature map are independent of each other and follow a many-to-one pipeline. They first determine a small local region in input feature map, according to a target location in output feature map. Then, they reassemble multiple features in the small local region with importance weighting kernels into one as the output feature for the target location. However, during the downsampling process, resembling features inside small local region independently for each target location ignores the coexist of multiple tiny lesions and limits ability to capture long-range dependencies for mutual activation enhancement between tiny lesions. These further lead to dilution or even vanishing of subtle activations caused by tiny lesions \cite{gao2019lip}, as illustrated in Fig. \ref{fig:M2M-M2One} (a). Once subtle lesion activations are vanished, it is almost impossible to recover them via upsampling, which consequently results in miss identification.

In this paper, we aim to address above issues raised by M2ORF operators within the context of tiny lesion segmentation. Basically, we pursue a RF operator that is capable of 1) preserving subtle activations caused by tiny lesions; 2) easily being integrated into existing CNN-based segmentation models. To this end, we propose a lightweight RF operator, termed as Many-to-Many Reassembly of Features (M2MRF), which considers the coexist of tiny lesions and enhances their subtle activations mutually via reassembling multiple features inside a large predefined region into multiple target features simultaneously (see Fig.\ref{fig:M2M-M2One}(b)). We demonstrate the effectiveness of our M2MRF on two public tiny lesion segmentation datasets, i.e. DDR \cite{li2019diagnostic} and IDRiD \cite{porwal2020idrid}. Experiments show that our M2MRF outperforms state-of-the-art RF operators. Particularly, compared with a strong segmentation baseline HRNetv2 \cite{wang2020deep}, our M2MRF exhibits significant improvements with negligible increase of parameters and inference time. 

The rest of this paper is organised as follows. The most related works are briefly reviewed in Section \ref{sec:related work}, and our proposed M2MRF and its application on tiny lesion segmentation are described in Section \ref{sec:method}. Section \ref{sec:exp} presents the experiments and analysis. The conclusion is presented in Section \ref{sec:conclusion}.

\section{Related Work}
\label{sec:related work}
\textbf{Feature reassembly operators in deep networks.} In existing deep networks \cite{long2015fully} \cite{ ronneberger2015u}   \cite{ chen2017deeplab} \cite{ chen2018encoder}  \cite{wang2020deep} \cite{sun2019deep} \cite{ noh2015learning} \cite{badrinarayanan2017segnet}, strided max-pooling \cite{lecun1990handwritten} and strided convolution are two widely-adopted operators for feature downsampling. Bilinear interpolation, deconvolution and unpooling\cite{zeiler2014visualizing} \cite{zeiler2011adaptive} are widely-adopted for feature upsampling. The general idea of these operators is to generate a feature vector for each target location via reassembling multiple features inside a predefined region with importance kernels. Particularly, importance kernels for strided max-pooling, unpooling and bilinear interpolation are hand-crafted and feature maps are processed channel-by-channel efficiently. However, they ignore the context dependencies across channels and the diversity of local patterns. Instead, the importance kernels for strided convolution and deconvolution are learned. Their dimension depends on the input features, which is always high in CNNs. This makes the computation burden of reassembly heavy when large importance kernels are used. Thus it is difficult to reassemble features from a large region.

Recently, novel ideas about learning-based RF operators emerge. LIP \cite{gao2019lip} learns adaptive importance kernels based on inputs to enhance the discriminative features. In CARAFE \cite{wang2019carafe} and CARAFE++ \cite{wang2020carafeplus}, content-aware kernels for each target position is learned according to input features for feature reassembly. Similarly, IndexNet \cite{lu2019indices} \cite{lu2020index} learns importance kernels from feature encoder to guide the feature reassembly in both feature encoder and decoder. Instead of reassembling features inside a predefined region, deformable RoI pooling \cite{dai2017deformable} reassembles features in an adaptive region which is learned from the input features. To model the affinity information, A$^2$U \cite{dai2021learning} is proposed to learn importance kernels according to second-order features for feature reassembly. However, these learning-based operators still follow the paradigm of many-to-one feature reassembly and are prone to dilute or even vanish features of tiny lesions.

\textbf{Tiny lesion segmentation in fundus images.} The development originated from one-type lesion segmentation with either hand-crafted features \cite{zhang2014exudate, liu2017location, du2020automatic} or deep features \cite{liu2021dsm} and has been evolved to multi-type lesion segmentation. For example, in \cite{li2019diagnostic}, PSPNet \cite{zhao2017pyramid} and DeeplabV3+ \cite{chen2018encoder} are directly fine-tuned for multi-type lesion segmentation. Guo et al. \cite{guo2019seg} develop a network named L-Seg on the top of FCN \cite{long2015fully}. More recently, multi-task learning frameworks \cite{foo2020multi}\cite{wang2021deep} are proposed to exploit the correlation among tasks of lesion segmentation and others to boost performance of all tasks. All of those methods adopt many-to-one RF operators during feature encoding and decoding and are prone to ignore tiny lesions. In \cite{sun2021lesion}, specially designed lesion filters are proposed to discover lesions and suspicious lesion regions are detected. However, this work is unable to provide fine lesion regions and predict lesion classes. One thing worth to mention is that the public datasets with pixel-level annotations also updates from one-type lesion segmentation e.g. E-ophtha EX \cite{decenciere2013teleophta} and e-ophtha MA \cite{decenciere2013teleophta} to multi-type lesion segmentation e.g. IDRiD \cite{porwal2020idrid} and DDR \cite{li2019diagnostic}.

\begin{figure}[tbh!]
	\begin{center}		
		\includegraphics[width=0.47\textwidth]{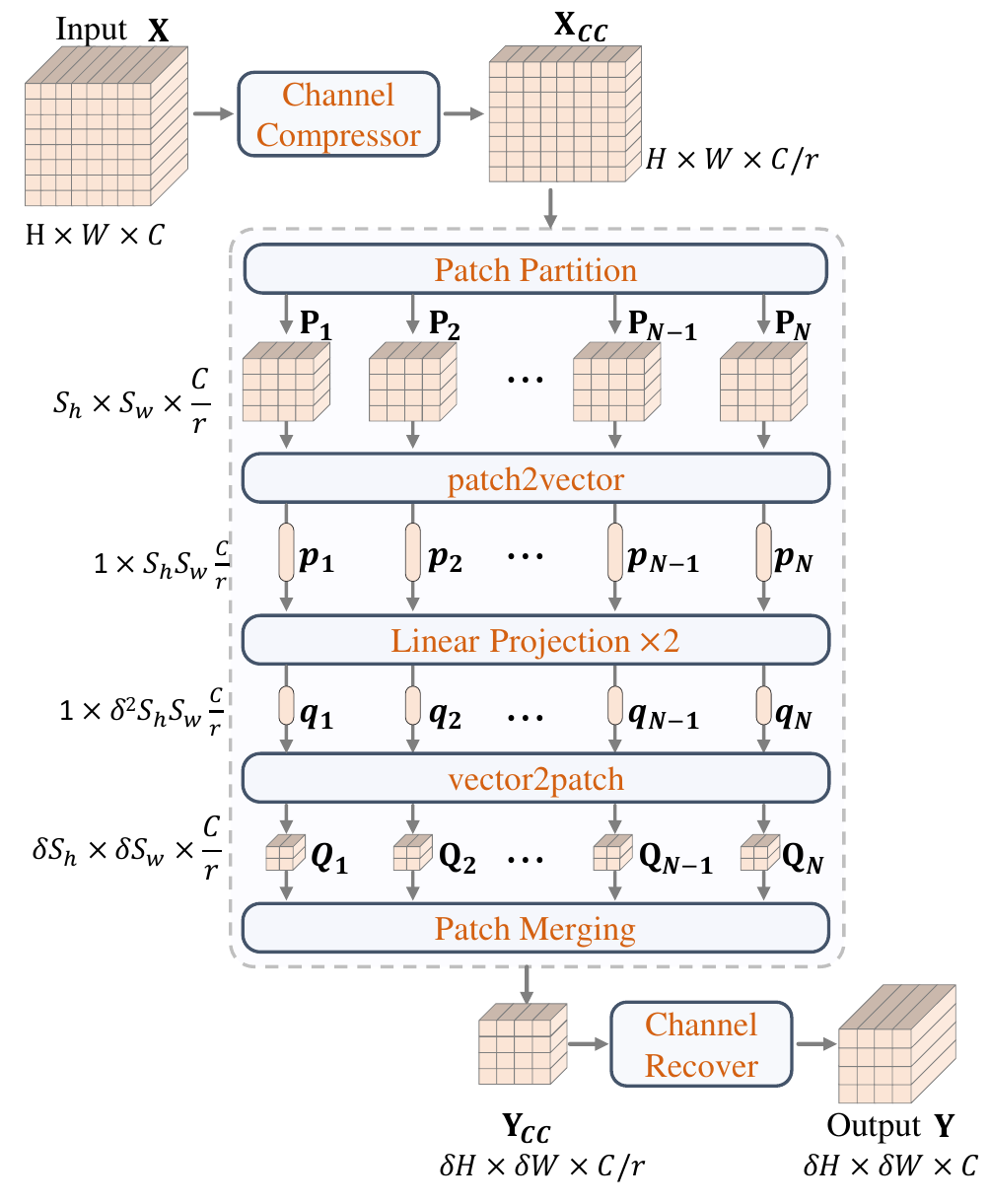}	
	\end{center}
	\caption{An overview of our Many-to-Many Reassembly of Features (M2MRF). $\{\mathbf{P}_l\}_{l=1}^L$ and $\{\mathbf{Q}_l\}_{l=1}^L$ are feature patches of size $ S_h \times S_w \times \frac{C}{r}$ and $\delta S_h \times \delta S_w \times \frac{C}{r}$ respectively, where $L(=\lceil H/S_h \rceil \cdot \lceil W/S_w \rceil)$ is the number of patches. $\{\mathbf{p}_l\}_{l=1}^L$ and $\{\mathbf{q}_l\}_{l=1}^L$ are feature vectors. In this figure, a feature map of size $H \times W \times C$ is downsampled by a factor of $\delta(=1/2)$.  (Best view in colour)}
	\label{fig:M2MRF-framework}
\end{figure}

\textbf{Deep semantic segmentation.} The development can be mainly classified into three veins. The first vein focuses on how to produce and aggregate multi-scale representations. FCN \cite{long2015fully} provides a natural solution which reuses middle-level features to compensate for spatial details in high-level features. UNet \cite{ronneberger2015u} propagates low-level information to high-levels via skip connections, which also inspires many variants specially for medical image segmentation such as VNet \cite{Diceloss}, UNet++ \cite{zhou2019unetpp}, MNet \cite{fu2018joint} and SAN \cite{liu2019spatial}. In \cite{li2020gated}, gated fully fusion  is proposed to selectively fuse multi-level features. The second vein focuses on high-resolution representation learning. For example, in \cite{ronneberger2015u,noh2015learning,badrinarayanan2017segnet}, an encoder-decoder architectural style is adopted to gradually recover feature resolution while HRNet \cite{sun2019deep, wang2020deep} is specially designed to learn high-resolution representations. The third vein introduces attention mechanism and variant modules such as DANet \cite{fu2019dual}, DRANet \cite{fu2020scene}, CCNet \cite{huang2019ccnet}, RecoNet \cite{chen2020tensor}, HANet \cite{choi2020cars} and DNL \cite{yin2020disentangled} are developed to explore long range dependencies. More recently, vision transformers such as Swin \cite{liu2021swin}, Twins \cite{chu2021Twins} and CSWin \cite{dong2021cswin} are developed for segmentation. However, those methods are designed for segmenting objects in proper size rather than tiny size.

\section{Method}
\label{sec:method}
\color{black}
In this section, we first give a simple analysis for M2ORF operators and then detail our M2MRF. Finally we take HRNetV2 \cite{wang2020deep} as an example and present how to integrate our M2MRF into CNN-architectures for tiny lesion segmentation.

\subsection{Analysis for M2ORF Operators}
Given the input feature map $\mathbf{X} \in \mathcal{R}^{H\times W \times C}$ and sample rate $\delta$ where $\delta > 0$, the goal of feature reassembly is to generate target feature map $\mathbf{Y}\in \mathcal{R}^{\lfloor\delta H\rfloor\times\lfloor\delta W\rfloor \times C}$ via finding a function mapping $\Phi$ parametrised by importance kernels $\mathbf{W}$:

\begin{equation}
	\label{eqn:framework}
	\mathbf{Y} = \Phi(\mathbf{X}; \mathbf{W})\;.
\end{equation}
Here $\delta<1$ for downsampling and $\delta > 1$ for upsampling. 

To make the feature reassembly computational efficient, most existing methods degrade it to a many-to-one local sampling problem, i.e. reassembling multiple features in a predefined local region to one target feature. Specifically, for any target feature $\mathbf{y}\in \mathcal{R}^C$ at location $(i',j')$ in $\mathbf{Y}$, most existing methods assume that there is a corresponding source feature $\mathbf{x}\in \mathcal{R}^{C}$ at location $(i, j)$ in $\mathbf{X}$, where $i =\lfloor i'/\delta \rfloor$ and $j = \lfloor j'/ \delta \rfloor$. They follow three steps to obtain $\mathbf{y}$ : (1) setting/learning a local region $\Omega_{\mathbf{x}}$ according to $\mathbf{x}\in \mathcal{R}^C$; (2) with $\Omega_{\mathbf{x}}$, setting/learning corresponding importance kernels $\mathbf{W}_{i',j'}$; (3) obtaining $\mathbf{y}$ via
\begin{equation}
	\label{eqn:singleframework}
	\mathbf{y} = \Phi(\mathbf{P}_{\Omega_{\mathbf{x}}}; \mathbf{W}_{i',j'})\;,
\end{equation}
where $\mathbf{P}_{\Omega_{\mathbf{x}}}$ denotes features in $\Omega_{\mathbf{x}}$.

Obviously, M2ORF operators assume that each target feature in $\mathbf{Y}$ is independent and ignores the coexist of multiple lesions. Additionally, considering the computational efficiency, previous RF operators such as strided convolution, deconvolution, CARAFE++ \cite{wang2020carafeplus} and A$^2$U \cite{dai2021learning} usually consider a small size $\Omega_{\mathbf{x}}$, which makes them fail to exploit long-range dependencies. As a result, activations caused by tiny lesions are prone to be diluted or even vanished. To alleviate the dilution of tiny lesion activations, one natural solution is to discard the assumption of M2ORF and generate multiple features for multiple target locations with features inside a large local region simultaneously. To this end, we propose many-to-many reassembly of features (M2MRF).

\subsection{Many-to-Many Reassembly of Features}
\textbf{Module overview.} Fig. \ref{fig:M2MRF-framework} illustrates an overview of reassembling the input feature map $\mathbf{X}$ of size $H\times W \times C$ with our M2MRF operator with sample rate $\delta$ to feature map $\mathbf{Y}$ of size $\lfloor\delta H\rfloor \times \lfloor \delta W \rfloor \times C$. First, a channel compressor is performed on $\mathbf{X}$ to reduce its channel from $C$ to $\frac{C}{r}$ for computational efficiency. We denote the output as $\mathbf{X}_{CC}$. Then we partition $\mathbf{X}_{CC}$ into feature patches $\{\mathbf{P}_l\}_{l=1}^L$ of size $S_h\times S_w \times \frac{C}{r}$, where $L = \lfloor H/S_h \rfloor  \cdot \lfloor W/S_w \rfloor$. Our proposed M2MRF is performed on each feature patch and outputs $\{\mathbf{Q}_l\}_{l=1}^L$ of size $\lfloor\delta H\rfloor \times \lfloor \delta W \rfloor \times \frac{C}{r}$. Thereafter, those patches are merged into feature map $\mathbf{Y}_{CC}$ of size $\lfloor\delta H\rfloor \times \lfloor \delta W \rfloor \times \frac{C}{r}$. Finally, we recover the feature channel to $C$ via channel recover. For channel compressor and recover, we simply implement them with a $1\times 1$ regular convolution layer.

\textbf{M2MRF.} With a local feature patch $\mathbf{P}\in \{\mathbf{P}_l\}_{l=1}^L$, the goal is to generate $\mathbf{Q}$ of size $\lfloor \delta S_h \rfloor \times \lfloor \delta S_w \rfloor \times \frac{C}{r}$:
\begin{equation}
	\label{eqn:m2mframework}
	\mathbf{Q} = \Phi(\mathbf{P}; \mathbf{W}_{patch})\;.
\end{equation}
Here we let $M=\lfloor \delta S_h \rfloor \times \lfloor \delta S_w \rfloor, N =  S_h\times S_w$, and treat this task as generating $M$ features $\mathbf{Q}=\{\mathbf{y}_m\}_{m=1}^{M}$ from $ N $ source features $\mathbf{P} = \{\mathbf{x}_n\}_{n=1}^{N}$ where $\mathbf{y}_m$, $\mathbf{x}_n \in \mathcal{R}^{1\times \frac{C}{r}}$. To achieve this, one option is to adopt linear projection, thus Eq.(\ref{eqn:m2mframework}) can be expressed as:
\begin{equation}
	\label{eqn:m2mframework_hd}
	[\mathbf{y}_1, \cdots, \mathbf{y}_{M}] = [\mathbf{x}_1, \cdots, \mathbf{x}_{ N}][\mathbf{W}_1, \cdots, \mathbf{W}_{M}]\;,
\end{equation}
where $\mathbf{W}_1,\cdots, \mathbf{W}_{M}$ are parameters of size $\frac{NC}{r} \times \frac{C}{r}$ to be learned. Therefore we have $\mathbf{W}_{patch}=[\mathbf{W}_1,\cdots, \mathbf{W}_{M}]$ whose size is $\frac{NC}{r} \times \frac{MC}{r}$. For simplicity, we denote $\mathbf{p} = [\mathbf{x}_1, \cdots, \mathbf{x}_{ N}],\mathbf{q} = [\mathbf{y}_1, \cdots, \mathbf{y}_{M}]$, and rewrite Eq.(\ref{eqn:m2mframework_hd}) as:

\begin{equation}
	\label{eqn:m2mframework_hd_vector}
	\mathbf{q} = \mathbf{p}\mathbf{W}_{patch}\;.
\end{equation}

To make use of long-range dependencies, $S_hS_w$ is required to be large. Accordingly, $N$ and $M$ are large and $\mathbf{W}_{patch}$ would be a large matrix. On one hand, it is always difficult to optimise such a large matrix. On the other hand, storing a large matrix results in high memory consumption. To reduce the number of parameters to be learned, we decompose $\mathbf{W}_{patch}$ to two small matrices via a two-layer linear projections. Thus Eq. (\ref{eqn:m2mframework_hd}) can be rewrote as:
\begin{equation}
	\label{eqn:mlp}
	\mathbf{q}= (\mathbf{p}\mathbf{W}_{patch}')\mathbf{W}_{patch}''\;,
\end{equation}
where $\mathbf{W}_{p}'\in \mathcal{R}^{\frac{NC}{r} \times \frac{NC}{\alpha r}}$ and $\mathbf{W}_{p}''\in \mathcal{R}^{\frac{NC}{\alpha r} \times \frac{MC}{r}}$ are parameters in the two linear projections, and $\alpha\geq 1$ such that the matrix dimension of $\mathbf{W}_{patch}'$ and $\mathbf{W}_{patch}''$ is less than $\mathbf{W}_{patch}$ far away.

\begin{figure*}[bth!]
	\begin{center}		
		\includegraphics[width=0.85\textwidth]{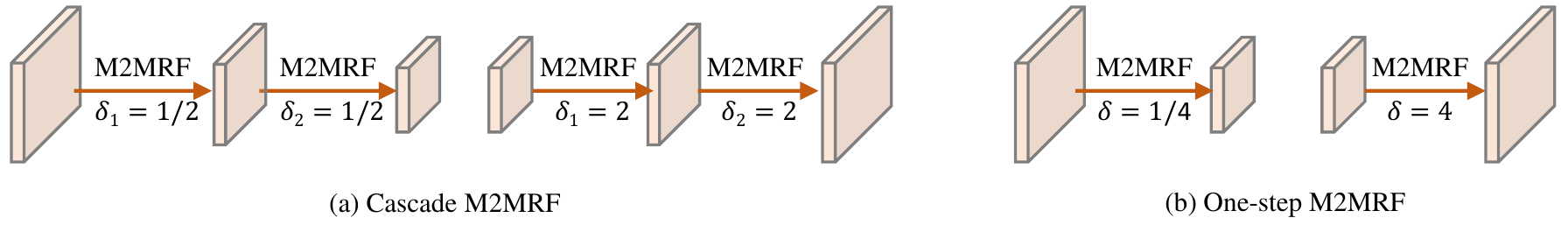}	
	\end{center}
	\caption{Two variants of our M2MRF. Here we take $\delta=4$ for upsampling and $\delta=1/4$ for downsampling as examples.}
	\label{fig:M2RFInvariants}
\end{figure*}

\subsection{M2MRF for Tiny Lesion Segmentation}
As HRNetV2 \cite{wang2020deep} is originally designed for high-resolution representation learning and has achieved state-of-the-art performance on semantic segmentation task, we adopt it as the baseline for tiny lesion segmentation. In HRNetV2 \cite{wang2020deep}, RF operators involve in the fusion module to exchange information across multi-resolution feature maps. Strided convolution with stride of 2 is adopted for downsampling and bilinear interpolation for upsampling. Repeating $t$ times of the strided convolution, the feature resolution decreases to $1/2^t$. To build M2MRF variants, we replace the repeated strided convolution and bilinear interpolation with our M2MRF.

%, in which the sample rate $\delta$ is set to $1/2$ and $2$ respectively. We name this modified HRNetV2 architecture as \textbf{M$^2$RF-HRNetV2}. 
We propose two options to replace the repeated $t$ layers of strided convolution with our M2MRF. One is to replace each strided convolution layer with our proposed M2MRF with $\delta=1/2$ to gradually decrease the feature resolution. We term it as cascade M2MRF. The other is to directly set  $\delta$ to $1/2^t$ in M2MRF to replace the entirety of $t$ layers of strided convolution. We term it as one-step M2MRF. Similarly, there are also two options to replace bilinear interpolation layer with scale factor $2^t$. One is to cascade M2MRF with $\delta=2$ to gradually increase the feature resolution to $2^t$. The other is one-step M2MRF with $\delta=2^t$ which directly increases the feature resolution to $2^t$. Taking $t=2$ for example, Fig.\ref{fig:M2RFInvariants} illustrates the cascade M2MRF and one-step M2MRF for downsampling and upsampling respectively.

With high-resolution feature maps by variants of HRNetV2 equipped with our M2MRFs, we attach multiple binary classifiers on them to obtain probability maps for each lesion class. Considering the extremely class imbalance in tiny lesion segmentation, we adopt Dice loss \cite{Diceloss} to train M2MRF-HRNetV2 as well as the baseline.
%\begin{equation}
%	\mathcal{Loss} = \sum_{l=1}^L \frac{1}{L}(1-Dice(\hat{Y}_l, G_l))\;,
%\end{equation}
%where $L$ is the total classes of tiny lesions, and $\hat{Y}_l$ is the predicted probability map of the $l$-th class and $G_l$ is ground truth mask, and $Dice$ is:
%\begin{equation}
%Dice(\hat{Y}_l, G_l) = \frac{\sum_{i=1}^{HW}\hat{Y}_{l,i}G{l,i}}{\sum_{i=1}^{HW}\hat{Y}_{l,i}^2 +\sum_{i=1}^{HW} G_{l,i}^2}
%\end{equation}

\section{Experiments and Discussions}
\label{sec:exp}
To validate the effectiveness of our M2MRF module on tiny lesion segmentation, we carry out experiments on two public datasets i.e. DDR \cite{li2019diagnostic} and IDRiD \cite{porwal2020idrid}. For evaluation metrics, we follow \cite{porwal2020idrid} and utilize Area Under Precision-Recall curve (AUPR). We also report mean F-score (mF) and mean IoU (mIoU) as they are widely adopted for segmentation evaluation.

\subsection{Implementation Details} 
\subsubsection{Datasets and augmentation}  DDR \cite{li2019diagnostic} contains 757 colour fundus images of size ranging from $1088\times 1920$ to $3456\times 5184$, among which 383 for training, 149 for validation and $225$ for testing. In DDR \cite{li2019diagnostic}, 24154, 13035, 1354 and 10563 connected regions are annotated by experts as EXs, HEs, SEs and MAs respectively. To our best knowledge, DDR \cite{li2019diagnostic} is the largest dataset for lesion segmentation in fundus images. IDRiD \cite{porwal2020idrid} contains 81 colour fundus images of size $4288\times 2848$ among which 54 for training and 27 for testing. It is provided by a grand challenge on “Diabetic Retinopathy – Segmentation and Grading” in 2018. In IDRiD \cite{porwal2020idrid}, 11716, 1903, 150 and 3505 connected regions are annotated by experts as EXs, HEs, SEs and MAs respectively. 

Before feeding images into models, we follow \cite{wang2021deep} and downsample images in DDR \cite{li2019diagnostic} such that the long side is 1024. Thereafter, zero padding is used on short side to enlarge its length to 1024. Following \cite{liu2021dsm, guo2019seg}, images in IDRiD \cite{porwal2020idrid} are resized to $1440\times 960$. Three tricks are adopted to augment the training data: multi-scale (0.5-2.0), rotation (90$^\circ$, 180$^\circ$ and 270$^\circ$) and flipping(horizontal and vertical).

\subsubsection{Experimental setting} M2MRF-HRNetV2 is built on the top of implementation of HRNetV2 \cite{wang2020deep} provided by MMSegmentation \cite{mmseg2020}. We initialize parameters associated with both M2MRF and dense classification layers with Gaussian distribution with zeros mean and standard deviation of 0.01 and the rest with the pre-trained model on ImageNet \cite{krizhevsky2012imagenet}. SGD is used for parameters optimisation. Hyper-parameters include: initial learning rate (0.01 poly policy with power of 0.9), weight decay (0.0005), momentum (0.9), batch size (4) and iteration epoch (60k on DDR and 40k on IDRiD).

\subsection{Results on DDR and Analysis}
\subsubsection{Ablation study for M2MRF}
There are four hyper-parameters in our M2MRF, i.e. the patch size $S_h$ and $S_w$, $r$ in channel compressor, $\alpha$ in Eq.(\ref{eqn:mlp}). For $S_h$ and $S_w$, we directly let them equal and conduct experiments with setting $\{4, 8, 16\}$. For $r$ and $\alpha$, we conduct experiments with setting $r$ to $\{2, 4, 8\}$ and $\alpha$ to $\{32, 64, 128\}$ and replacing the strided convolution layer in HRNetV2 \cite{wang2020deep} with our cascade M2MRF and bilinear interpolation layer in HRNetV2 \cite{wang2020deep} with one-step M2MRF. Performance on DDR \cite{li2019diagnostic} validation set is listed in Table \ref{tab:AblationHyperP}, from which we can see that $S_h = S_w = 8$, $r=4$ and $\alpha=64$ yield best performance. In what followed, except for extra illustration, $S_h = S_w = 8$, $r=4$ and $\alpha=64$ are default setting.

\begin{table}[tbh!]
	\begin{center}
		\begin{tabular}{|ccc|ccc|}
			\hline
			$S_h, S_w$ & $r$ & $\alpha$ & mAUPR & mF & mIoU\\\hline	
			4	&	4	&	64	&	60.93 	&	59.77 	&	43.33 	\\	
			\textbf{8}	&	\textbf{4}	&	\textbf{64}	&	\textbf{61.49} 	&	\textbf{60.26} 	&	\textbf{43.80} 	\\	
			16	&	4	&	64	&	60.69 	&	59.42 	&	42.93 	\\	\hline
			8	&	2	&	64	&	61.13 	&	59.89 	&	43.45 	\\	
			\textbf{8}	&	\textbf{4}	&	\textbf{64}	&	\textbf{61.49} 	&	\textbf{60.26} 	&	\textbf{43.80} 	\\
			8	&	8	&	64	&	61.43 	&	60.18 	&	43.78 	\\	\hline
			8	&	4	&	32	&	60.34 	&	59.24 	&	42.71 	\\	
			\textbf{8}	&	\textbf{4}	&	\textbf{64}	&	\textbf{61.49} 	&	\textbf{60.26} 	&	\textbf{43.80} 	\\
			8	&	4	&	128	&	61.31 	&	60.06 	&	43.63 	\\	\hline			
		\end{tabular}
	\end{center}
	\caption{M2MRF(cascade/one-step) with different settings on DDR validation set \cite{li2019diagnostic}. Results are averaged over three repetitions.}
	% Empirically, $r_1=4, r_2=64$ yields best performance.
	\label{tab:AblationHyperP}
\end{table}

\begin{table}[tbh!]	
	\begin{center}
		\begin{tabular}{|c|l|ccc|cc|}
			\hline
			\multicolumn{2}{|c|}{operators}		&	mAUPR & mF	&	mIoU	&	Param(M)	&	FPS	\\\hline
			\multirow{6}{*}{DS}
			& \textit{StrideConv}	&	45.21 	&43.95 	&	28.84 	&	65.85 	&	\underline{11.31}	\\
			&MaxPool		&	45.97	&44.28 &	29.17	&	\textbf{59.45} 	&	\textbf{11.88} 	\\
			&LIP \cite{gao2019lip}		&	43.14 	&40.68 &	26.29 	&	75.44 	&	6.80 	\\
			&CARAFE++ \cite{wang2020carafeplus} &	46.27 &43.51 &	28.51 & 66.30 	&	9.24 \\ 	\cline{2-7}												
			&one-step M2MRF	&	\textbf{49.41} 	&	\textbf{45.68} 	&	\textbf{30.36} 
			&	\underline{60.75} 	&	9.83 	 	\\ 
			& cascade M2MRF	&	\underline{48.92} 	&	\underline{44.99} 	&	\underline{29.71} 	&	61.12 	&	9.28 	\\ \hline
			
			\multirow{5}{*}{US} & \textit{Bilinear}	&	45.21 	& 43.95 &	28.84 	&	\textbf{65.85} 	&	\textbf{11.31}\\			
			& Deconv	&	\underline{46.24} 	& \underline{44.59} &	\underline{29.37} 	&	73.12 	&	\underline{10.86}  	\\			
			& CARAFE++\cite{wang2020carafeplus}	&	\textbf{46.68} 	& \textbf{45.32} &	\textbf{29.97} 	& \underline{72.12} &	10.26 \\	\cline{2-7}		
			& one-step M2MRF	& 45.44 & 	44.03 & 28.88  	&	75.03 &	9.69 \\
			& cascade M2MRF	&45.37 	& 44.46 & 29.27 &	72.25	& 9.16	\\\hline	
			%\multirow{9}{*}{Paired (DS/UP)} & \textit{StrideConv/Bilinear}	&	45.21 	&43.95	&	28.84 	&	65.85 	&	11.31
			%\\		
			%& CARAFE++\cite{wang2020carafeplus}	&	47.64 	& 44.80  &	29.54 	&	72.57 	&	8.69 	 \\	
			
			%&MaxPool/Unpooling	&	48.81	& \underline{46.17}  &	\underline{30.73}	&	\textbf{59.45}  	&	%\underline{11.62}	\\
			
			%&IndexNet\cite{lu2020index}	&	48.06 	& 45.64 
			%&	30.28 	&	70.33 	&	10.75 	 \\				
			
			%	& A$^2$U\cite{dai2021learning}	&	45.89 	& 44.44 &	29.27 	&	66.51 	&	3.97 		\\\cline{2-7}									
			%& M2MRF-A	&	\textbf{49.94} & \textbf{46.60} &	\textbf{31.16} 	&	69.93 	&	9.21 	\\
			%& M2MRF-B	&	\underline{49.42} &	45.77 & 30.41 	&	67.15 	&	8.54 	\\
			%& M2MRF-C	&	48.94 	&45.40&	30.09 	&	70.30 	&	8.43 		\\
			%& M2MRF-D	&	49.25 	&45.57 &	30.27 	&	67.52 	&	8.01 	 	\\\hline
		\end{tabular}
	\end{center}
	\caption{Comparing with different downsampling (DS) and upsampling (US) operators on DDR \cite{li2019diagnostic} testing set. StrideConv as downsampling operator and bilinear interpolation as upsampling operator are the default set in vanilla HRNetV2 \cite{wang2020deep}.  The Best and Second Best results are highlighted in boldface and underlined respectively. Results are averaged over three repetitions.}
	\label{tab:DSComparisonDDR}
\end{table}

\begin{figure*}[tbh!]
	\begin{center}		
		\includegraphics[width=0.9\textwidth]{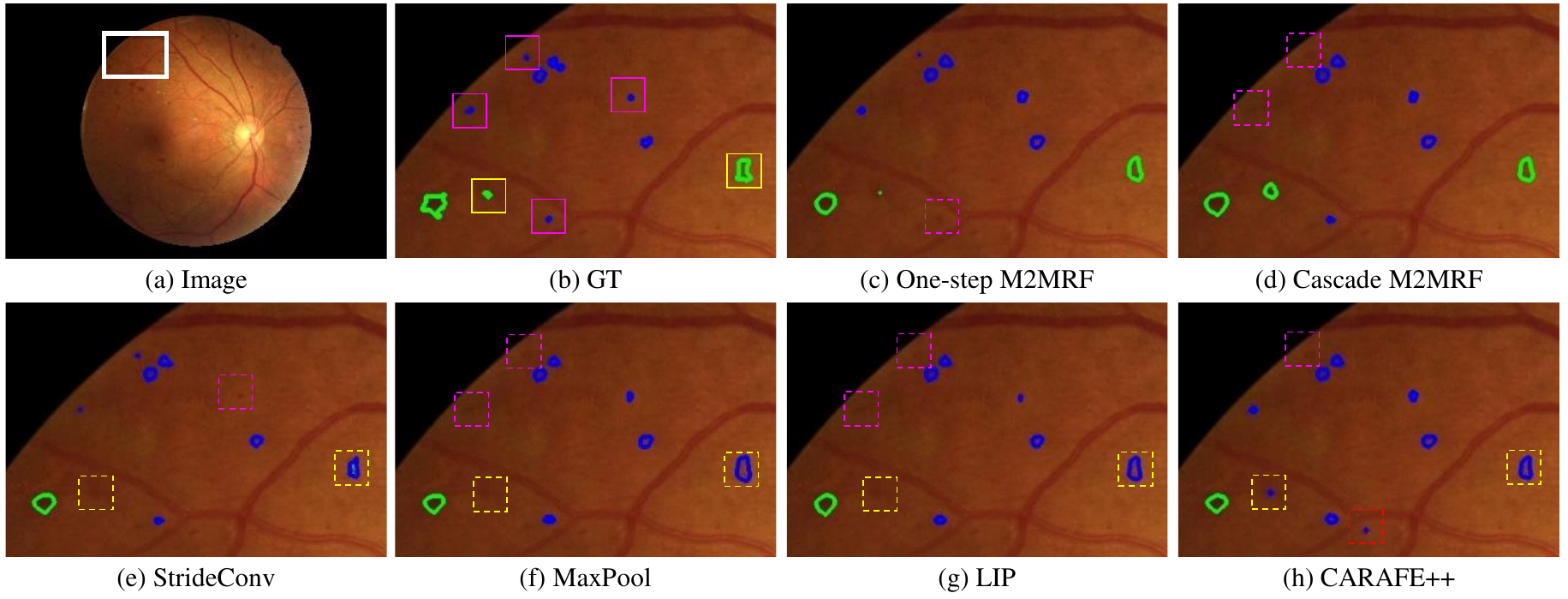}	
	\end{center}
	\caption{Visualization results on DDR test set \cite{li2019diagnostic} by variants of HRNetV2 with different downsampling operators, i.e., our one-step M2MRF, cascade M2MRF, StrideConv \cite{wang2020deep}, MaxPool, LIP \cite{gao2019lip} and CARAFE++ \cite{wang2020carafeplus}. Regions delineated in green and blue are HEs and MAs by experts or segmentation approaches. As red lesions, MAs and tiny HEs have deceptive appearances and both exhibit as dark red dots. In GT map (b), challenge lesions are marked with solid boxes. In visualised segmentation result maps (c-h), wrong and miss identifications are marked with dotted boxes.}
	\label{fig:ddr-ds}
\end{figure*}

\begin{figure*}[tbh!]
	\begin{center}		
		\includegraphics[width=0.9\textwidth]{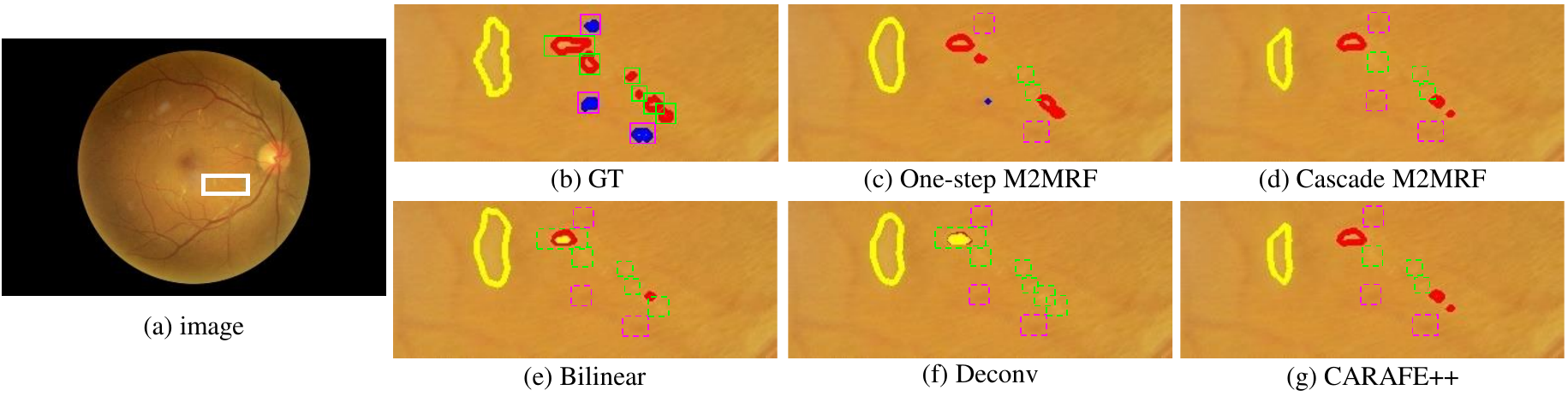}	
	\end{center}
	\caption{Visualization results on DDR test set \cite{li2019diagnostic} by variants of HRNetV2 with different upsampling operators, i.e., our one-step M2MRF, cascade M2MRF, bilinear interpolation \cite{wang2020deep}, deconvolution and CARAFE++ \cite{wang2020carafeplus}. Regions delineated in blue, red and yellow are MAs, EXs and SEs by experts or segmentation approaches. In GT map (b), challenge lesions are marked with solid boxes. In visualised segmentation result maps (c-g), wrong and miss identifications are marked with dotted boxes.}
	\label{fig:ddr-us}
\end{figure*}

\subsubsection{Comparison to other RF operators} Here we verify the effectiveness of our proposed M2MRF on DDR \cite{li2019diagnostic} testing set. In what follows, we first compare our M2MRF with downsampling and upsampling operators, and then RF operators that can/must be used as pairs.

\textbf{Comparison to downsampling and upsampling operators}. For downsampling, we replace strided convolution operator (StrideConv) in HRNetV2 \cite{wang2020deep} with three optional operators, {i.e.}, strided max-pooling (MaxPool), LIP \cite{gao2019lip} and CARAFE++ \cite{wang2020carafeplus}, and our two variants of M2MRF, {i.e.}, cascade M2MRF and one-step M2MRF, and make comparisons on DDR \cite{li2019diagnostic} test set. Among those operators, MaxPool is rule-based and parameter-free while rest are learning-based.

StrideConv is the default setting in vanilla HRNetV2 \cite{wang2020deep}, whose kernel size is set to $3\times 3$ and stride is set to 2. For MaxPool, $3\times 3$ max-pooling kernel with stride of 2 is adopted. For LIP \cite{gao2019lip} and CARAFE++ \cite{wang2020carafeplus}, the suggested settings by original papers are adopted. We replace StrideConv in HRNetV2 \cite{wang2020deep} with those compared downsampling operators and apply to tiny lesion segmentation. Table \ref{tab:DSComparisonDDR} reports their performances. We can see that: (1) Rule-based downsampling operator, i.e., MaxPool, achieves better performance than StrideConv slightly in terms of mAUPR, mF and mIoU consistently; (2) LIP \cite{gao2019lip} seriously degrades the segmentation performance. Possible reason is that LIP \cite{gao2019lip} adopts instance normalisation to normalise the importance kernels globally over entire feature map of each channel, which may heavily dilute subtle activations caused by tiny lesions and result in miss identification on tiny lesions; (3) CARAFE++ \cite{wang2020carafeplus} improves mAUPR by 1.06\% but slightly decreases mF by 0.44\% and mIoU by 0.33\%; (4) Our two variants of M2MRF improve mAUPR, mF and mIoU consistently. In detail,  cascade M2MRF improves mAUPR, mF and mIoU by 3.71\%, 1.04\% and 0.87\% respectively while one-step improves mAUPR, mF and mIoU by 4.20\%, 1.73\% and 1.52\% respectively.

We also list the parameters and inference speed of variants of HRNetV2 equipped with the compared downsampling operators in Table \ref{tab:DSComparisonDDR}. Comparing with baseline, our proposed M2MRFs have less parameters but scarifies slight inference speed. Comparing with LIP \cite{gao2019lip} and CARAFE++ \cite{wang2020carafeplus}, our proposed M2MRFs have less parameters and competitive inference speed. Obviously, HRNetV2 \cite{wang2020deep} equipped with MaxPool has least parameters and achieves fastest inference speed as MaxPool is rule-based and parameter-free.

\begin{table}[tbh!]	
	\begin{center}
		\begin{tabular}{|l|ccc|cc|}
			\hline
			RF operators		&	mAUPR	& mF &	mIoU	&	Param(M)	&	FPS	\\\hline
			StrideConv/Bilinear	&	45.21 	&43.95	&	28.84 	&	\underline{65.85} 	&	\underline{11.31}
			\\		
			CARAFE++\cite{wang2020carafeplus}	&	47.64 	& 44.80  &	29.54 	&	72.57 	&	8.69 	 \\	
			MaxPool/Unpooling	&	48.81	& \underline{46.17}  &	\underline{30.73}	&	\textbf{59.45}  	&	\textbf{11.62}	\\
			IndexNet\cite{lu2020index}	&	48.06 	& 45.64 
			&	30.28 	&	70.33 	&	10.75 	 \\						
			A$^2$U\cite{dai2021learning}	&	45.89 	& 44.44 &	29.27 	&	66.51 	&	3.97 		\\\hline
			M2MRF-A	&	\textbf{49.94} & \textbf{46.60} &	\textbf{31.16} 	&	69.93 	&	9.21 	\\
			M2MRF-B	&	\underline{49.42} &	45.77 & 30.41 	&	67.15 	&	8.54 	\\
			M2MRF-C	&	48.94 	&45.40&	30.09 	&	70.30 	&	8.43 		\\
			M2MRF-D	&	49.25 	&45.57 &	30.27 	&	67.52 	&	8.01 	 	\\\hline
		\end{tabular}
	\end{center}
	\caption{Comparing with different paired RF operators on DDR \cite{li2019diagnostic} testing set. M2MRF-A: (one-step/one-step), M2MRF-B:(one-step/cascade), M2MRF-C:(cascade/one-step), M2MRF-D: (cascade/cascade). The first row is the baseline i.e. vanilla HRNetV2 \cite{wang2020deep}. The Best and Second Best results are highlighted in boldface and underlined respectively. All results are averaged over three repetitions.}
	\label{tab:RFcomparison}
\end{table}

\begin{figure*}[tbh!]
	\begin{center}		
		\includegraphics[width=0.9\textwidth]{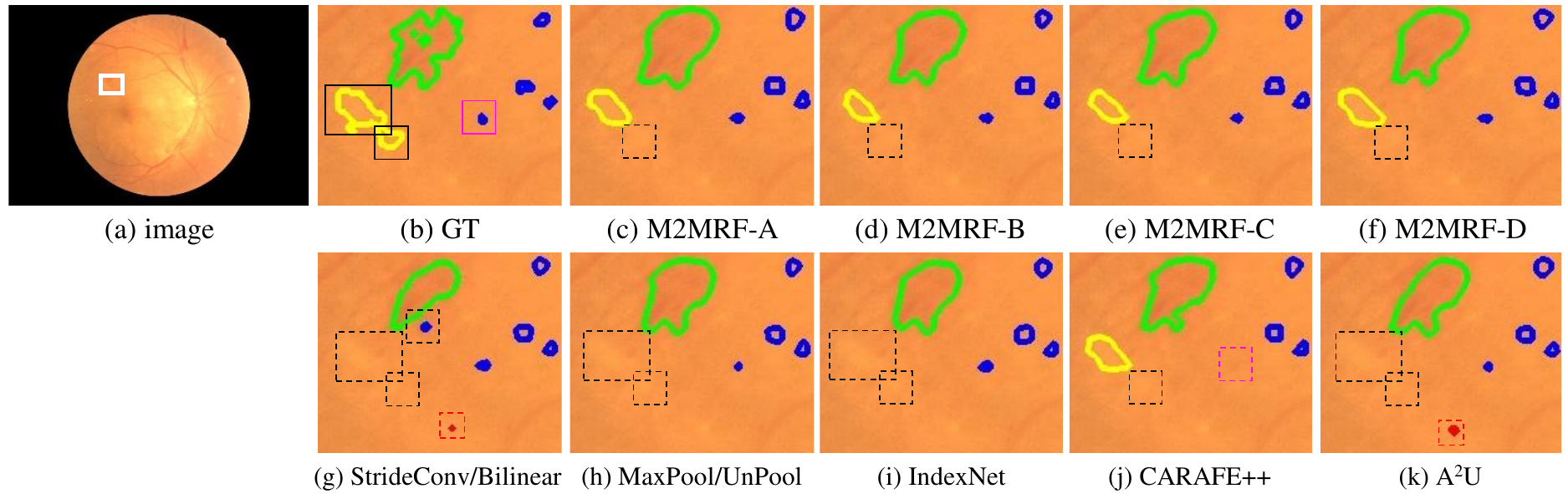}	
	\end{center}
	\caption{Visualization results on DDR test set \cite{li2019diagnostic} by variants of HRNetV2 with paired RF operators: M2MRF-A, M2MRF-B, M2MRF-C, M2MRF-D, StrideConv/bilinear, MaxPool/UnPool, IndexNet \cite{lu2020index}, CARAFE++ \cite{wang2020carafeplus} and A$^2$U \cite{dai2021learning}. Regions delineated in green, blue, red and yellow are HEs, MAs, EXs and SEs by experts or segmentation approaches. In GT map (b), challenge lesions are marked with solid boxes. In visualised segmentation result maps (c-k), wrong and miss identifications are marked with dotted boxes.}
	\label{fig:ddr-rf}
\end{figure*}

\begin{table*}[tbh!]
	\begin{center}
		\begin{tabular}{|l|c c c c|c|c c c c|c|c c c c|c|}
			\hline
			\multirow{2}{*}{Methods}&\multicolumn{5}{c|}{AUPR}  & \multicolumn{5}{c|}{F} & \multicolumn{5}{c|}{IoU} \\\cline{2-16}
			& EX & HE & SE & MA & mAUPR & EX & HE & SE & MA &  mF & EX & HE & SE & MA &  mIoU\\ \hline
			HED\cite{xie2015holistically}	&	61.40 	&	43.19 	&	46.68 	&	20.61 	&	42.97 	 	
			&	56.63 	&	42.61 	&	45.50 	&	22.43 	&	41.79 				&	39.50 	&	27.09 	&	29.46 	&	12.63 	&	27.17 \\
			
			DNL\cite{yin2020disentangled}	&	56.05 	&	47.81 	&	42.01 	&	14.71 	&	40.14 	
			&	53.36 	&	42.71 	&	40.40 	&	15.60 	&	38.02 	
			&	36.39 	&	27.15 	&	25.33 	&	8.46 	&	24.33 	
			\\
			
			Deeplabv3+\cite{chen2018encoder}	&	62.32 	&	40.79 	&	41.83 	&	24.39 	&	42.34 	
			&	58.59 	&	37.97 	&	41.83 	&	25.40 	&	40.95
			&	41.44 	&	23.44 	&	26.46 	&	14.55 	&	26.47 	
			\\
			
			PSPNet\cite{zhao2017pyramid}	&	57.04 	&	42.71 	&	42.32 	&	14.85 	&	39.23 	
			&	54.35 	&	39.37 	&	42.08 	&	16.09 	&	37.97 
			&	37.31 	&	24.51 	&	26.64 	&	8.75 	&	24.31 	
			\\
			
			SPNet\cite{hou2020strip}	&	44.10 	&	38.22 	&	32.93 	&	12.37 	&	31.91 	
			&	38.78 	&	24.13 	&	34.00 	&	13.74 	&	27.66
			&	24.19 	&	13.76 	&	20.55 	&	7.38 	&	16.47 	
			\\
			
			L-seg \cite{guo2019seg} & 55.46 & 35.86 & 26.48 & 10.52 & 32.08& -& -& -& -& - &-&-&-&-&- \\
			
			HRNetV2\cite{wang2020deep}	&	61.55 	&	45.68 	&	46.91 	&	26.70 	&	45.21 	
			&	58.98 	&	44.96 	&	44.86 	&	26.99 	&	43.95
			&	41.82	&	29.01 	&	28.94 	&	15.60 	&	28.84 	
			\\
			\hline
			Swin-base\cite{liu2021swin}	&	62.95 &	53.46 &	50.56 &	23.46 &	47.61 
			
			&60.12 &	\textbf{51.10} &	50.85 &	23.38 &	\underline{46.36} 
			
			& 	42.98 &	\textbf{34.42} &	34.15 &	13.24 &	\textbf{31.20} 
			
			\\	
			
			Twins-SVT-B \cite{chu2021Twins} &
			59.71 	&	49.96 	&	\underline{52.72} 	&	22.03 	&	46.11 	
			&	56.83 	&	45.04 	&	\textbf{53.19} 	&	21.54 	&	44.15
			&	39.70 	&	29.08 	&	\textbf{36.24} 	&	12.07 	&	29.28 	
			\\
			
			\hline				
			
			M2MRF-A&	\textbf{64.17} 	&	54.20 	&	\textbf{53.19} 	&	28.21 	
			&	\textbf{49.94} 	
			&	60.47 	&	46.18	&	\underline{52.10} 				&	27.67 &	\textbf{46.60} 
			&	43.35 	&	30.03 	&	\underline{35.22} 	&	16.06 &	\underline{31.16} 
			
			\\
			M2MRF-B&63.88 	&	\underline{55.47}	&	50.01 	&	28.33 	&	\underline{49.42} 	
			&	60.20 	&	\underline{46.81} 				&	48.58 	&	27.51 	&	45.77
			&	43.06 	&	\underline{30.56} 	&	32.08 	&	15.95 	&	30.41 
			
			\\			
			M2MRF-C&	63.59 	&	54.43 	&	49.35 	&	\underline{28.38} 	&	48.94 	
			&	\underline{60.62} 				&	45.16 	&	47.78 	&	\textbf{28.04} 			&	45.40 			
			&	\underline{43.49} 	&	29.17 	&	31.39 	&	\textbf{16.31} 	&	30.09 
			
			\\
			
			M2MRF-D&	\textbf{64.17} 	&	\underline{54.72} 	&	49.64 	&	\textbf{28.46} 	&	49.25 
			&\textbf{61.15} &	45.29 &	48.02 &	\underline{27.81} &	45.57
			&	\textbf{44.04} 	&	29.28 	&	31.60 	&	\underline{16.15} 	&	30.27 
			
			\\ \hline					
		\end{tabular}
	\end{center}
	\caption{Segmentation results on DDR test set \cite{li2019diagnostic}. M2MRF-A: (one-step/one-step), M2MRF-B:(one-step/cascade), M2MRF-C:(cascade/one-step), M2MRF-D: (cascade/cascade). The Best and Second Best results are highlighted in boldface and underlined respectively. All results are averaged over three repetitions.}
	\label{tab:DDR}
\end{table*}

To further illustrate the superiority of our M2MRFs as downsampling operators, we visualise segmentation results on MAs and HEs in Fig. \ref{fig:ddr-ds}. The reason of choosing these two types of lesions is that 
MAs and tiny HEs belong to red lesions and exhibit as dark red dots with similar appearances. Correctly identifying them requires that segmentation models are able to maintain subtle activations about them and exploit long-range dependencies to boost the discriminativeness of activations during downsampling. For clarity, we mark challenge MAs with solid magenta boxes and HEs with solid yellow boxes in GT map (see Fig. \ref{fig:ddr-ds}(b)) and miss and wrong identifications by variants of HRNetv2 \cite{wang2020deep} utilising different downsampling operators in Fig. \ref{fig:ddr-ds} (c-h) with dotted boxes. Obviously, our proposed cascade M2MRF and one-step M2MRF (see Fig. \ref{fig:ddr-ds}(c-d)) encounter less miss and wrong identifications than compared four downsampling operators as shown in Fig. \ref{fig:ddr-ds}(e-h). Particularly, both of our cascade M2MRF and one-step M2MRF are able to correctly distinguish between HEs and MAs while variants of HRNetv2 \cite{wang2020deep} with rest downsampling operators are prone to be deceived by their confusion appearances and wrongly identify HEs as MAs. Although CARAFE++ \cite{wang2020carafeplus} is able to maintain subtle activations about tiny lesions, it seems scarifies the feature discrimination and tends to cause wrong identifications.

For upsampling, we utilise strided convolution for downsampling and replace bilinear interpolation in HRNetV2 \cite{wang2020deep} with four optional upsampling operators, {i.e.}, deconvolution and CARAFE++ \cite{wang2020carafeplus}, our proposed cascade M2MRF and one-step M2MRF. Among them, bilinear interpolation is rule-based and parameter-free while the rest are learning-based. We compare their performance on DDR \cite{li2019diagnostic} test set. As reported in Table \ref{tab:DSComparisonDDR}, comparing with bilinear interpolation, our two variants of M2MRF achieve slightly better mAUPR, mF and mIoU but require extra parameters and inference time. From Table \ref{tab:DSComparisonDDR} we also can see that, different to CARAFE++ \cite{wang2020carafeplus} which works better as upsampling operator than downsample operator, our M2MRFs work better as downsampling operators. Fig. \ref{fig:ddr-us} shows visualised segmentation results. As exampled, it is challenge to well segment tiny lesions clustering together. Comparatively speaking, our one-step M2MRF, cascade M2MRF and CARAFE++ \cite{wang2020carafeplus} make less mistakes than bilinear interpolation and deconvolution.

%Different from CARAFE++ \cite{wang2020carafeplus} which works better when only used to replace the upsampling operators in HRNetv2 \cite{wang2020deep}, our M2MRFs work better when used for downsampling. 

%\begin{table}[tbh!]	
%	\begin{center}
%		\begin{tabular}{|l|ccc|cc|}
%			\hline
%			Upsample		&	mAUPR & mF	&	mIoU	&	Param(M)	&	FPS	\\\hline
%			Bilinear	&	45.21 	& 43.95 &	28.84 	&	65.85 	&	11.31\\			
%			Deconv	&	46.24 	&44.59 &	29.37 	&	73.12 	&	10.86  	\\			
%			CARAFE\cite{wang2019carafe}	&	46.68 	& 45.32 &	29.97 	& 72.12 &	10.26 	 		
%			\\\hline
%				M2MRF(cascad)	&45.37 	& 44.46 & 29.27 &	72.25	& 9.16	\\
%				M2MRF(one-step)	& 45.44 & 	44.03 & 28.88  	&	75.03 &	9.69 \\\hline
%		\end{tabular}
%	\end{center}
%	\caption{Comparing with different upsampling operators on DDR \cite{li2019diagnostic} testing set. The first row is the baseline i.e. vanilla HRNetV2 \cite{wang2020deep}. The Best and Second		Best results are highlighted in boldface and underlined respectively. Results are averaged over three repetitions.}
%	\label{tab:USComparisonDDR}
%\end{table}

% ( many-to-many feature reassembly module)

\textbf{Comparison to paired RF operators}. We have four different RF pairs when combining cascade M2MRF and one-step M2MRF for both downsampling and upsampling. We denote them by M2MRF-A (one-step M2MRF for both downsampling and upsampling), M2MRF-B (one-step M2MRF for downsampling and cascade M2MRF for upsampling), M2MRF-C (cascade M2MRF for downsampling and one-step M2MRF for upsampling), M2MRF-D (cascade M2MRF for both downsampling and upsampling). We compare them with five pairs of RF operators, {i.e.}, StrideConv/Bilinear, CARAFE++ \cite{wang2020carafeplus}, MaxPool/Unpooling, IndexNet \cite{lu2020index}  and A$^2$U \cite{dai2021learning} on DDR \cite{li2019diagnostic} test set. Among those RF operators, MaxPool/Unpooling and IndexNet \cite{lu2020index} have to utilise the indices generated during downsampling for upsampling. Table \ref{tab:RFcomparison} reports quantitative results. As listed, among four variants of our M2MRF, M2MRF-A achieves best in mAUPR, mF and mIoU. Comparing with vanilla HRNetV2 \cite{wang2020deep}, our M2MRF-A improves the mAUPR by a large margin from 45.21\% to 49.94\% and the mIoU from 28.84\% to 31.16\% with few extra parameters (65.85M$\rightarrow$69.93M) and slightly slower inference speed (11.31fps$\rightarrow$9.21fps). Comparing with existing RF operators, our M2MRF-A achieves best in both mAUPR and mIoU and the rest three variants achieve better mAUPR and competitive mIoU. Specifically, comparing with the recent learning-based RF operators IndexNet \cite{lu2020index} and CARAFE++ \cite{wang2020carafeplus}, our M2MRF-A and M2MRF-B achieve better performance in mAUPR and mIoU with less parameters and competitive inference speed. Comparing with A$^2$U \cite{dai2021learning}, our four variants of M2MRF outperform it on both mAPUR and mIoU with faster inference speed. Fig. \ref{fig:ddr-rf} shows visualised segmentation results, from which we can see that our four variants of M2MRF are able to preserve very subtle lesion activations and make less mistakes than those compared paired RFs.

\subsubsection{Comparison to state-of-the-art segmentation methods} We compare our M2MRF-HRNetV2 with nine state-of-the-art segmentation approaches as listed in Table \ref{tab:DDR}. The first seven methods are CNN-based and the rest two are transformer-based. Results by Swin-base \cite{liu2021swin} are obtained by fine-tuning the pre-trained model on ImageNet-22K with DDR training set \cite{li2019diagnostic} and the rest are obtained by fine-tuning the pre-trained models on ImageNet-1K \cite{krizhevsky2012imagenet}. As we can see that: (1) among  previous CNN-based segmentation methods, the vanilla HRNetV2 \cite{wang2020deep} achieves best performance with 45.21\% on mAUPR, 43.95\% on mF and 28.84\% on mIoU; (2) comparing with vanilla HRNetV2 \cite{wang2020deep}, the four variants of our M2MRF get better performance on the four lesion classes consistently in terms of mAUPR, mF and mIoU; (3) comparing with one of the most recent transformer-based methods, Swin-base \cite{liu2021swin}, our four variants of M2MRF achieve better on mAUPR by a large margin while our M2MRF-A achieves competitive performance on mF and mIoU; (4) comparing with another transform-based method Twins-SVT-B \cite{chu2021Twins}, the four variants of our M2MRF surpass it significantly and consistently on mAUPR, mF and mIoU.

%Fig. \ref{fig:ddr} shows more visualisation results of our M2MRF-HRNetV2 and the vanilla HRNetV2 \cite{wang2020deep} on DDR test set \cite{li2019diagnostic}. In the first example, we can see that our M2MRF(one-step/one-step)-HRNetV2 is able to segment out the two SEs (delineated in yellow) even though they are surrounded by HEs (delineated in green). On the contrary, the vanilla HRNetV2 \cite{wang2020deep} fails to well segment them out. In clinical, HEs and MAs belong to red lesions. The difference between small HEs and MAs is subtle. Thus our method wrongly classifies the small HEs as MAs. However, more worse, those small HEs are directly ignored by vanilla HRNetV2 and wrongly classified as background. In the second and third examples, tiny EXs pointed out by white arrows are correctly segmented out by our method while miss-classified by vanilla HRNetV2. 

%\begin{table}.
%	\label{tab:rst_eophtha}
%	\begin{center}
%		\begin{tabular}{l|c|c|c|r}
%			\hline
%			method & training & test & AUPR & IoU \\\hline	
%			Baseline & e-ophtha EX & e-ophtha EX &  52.79 & 35.64 \\			
%			M2MRF & e-ophtha EX & e-ophtha EX &   & \\		\hline		
%			Baseline & e-ophtha MA & e-ophtha MA & 45.42  & 29.82\\			
%			M2MRF & e-ophtha MA & e-ophtha MA &   & \\	\hline		
%			Baseline & IDRiD & e-ophtha EX & \textbf{66.29} & \textbf{47.39} \\
%			M2MRF & IDRiD & e-ophtha EX & 65.80& 45.80 \\ 		\hline		
%			M2MRF & IDRiD & e-ophtha MA &\textbf{45.27} &\textbf{29.52} \\\hline
%		\end{tabular}
%	\end{center}
%	\caption{Comparing with different feature reassembly methods on e-ophtha EX and e-ophtha MA.}
%\end{table}

\begin{table}[tbh!]	
	\begin{center}
		\begin{tabular}{|l|ccc|}
			\hline
			RF operators		&	mAUPR	& mF &	mIoU \\\hline
			StrideConv/Bilinear	&	65.01 	&	63.14 	&	47.52 	\\
			MaxPool/Bilinear	&	66.24 	&	65.01 	&	49.28 	\\
			LIP \cite{gao2019lip}/Bilinear		&	63.04 	&	61.50 	&	45.78 	\\
			StrideConv/Deconv	&	64.64 	&	62.86 	&	47.23 	\\
			MaxPool/Unpooling	&	66.17 	&	64.98 	&	49.22 	\\
			CARAFE++ \cite{wang2020carafeplus}			&	66.35 	&	64.67 	&	48.91 	\\
			IndexNet \cite{lu2020index}			&	65.77 	&	64.30 	&	48.74 	\\
			A$^2$U	\cite{dai2021learning}			&	64.93 	&	63.23 	&	47.52 	\\ \hline
			M2MRF-A				&	66.48 	&	64.89 	&	49.07 	\\			
			M2MRF-B				&	66.00 	&	64.45 	&	48.56 	\\
			M2MRF-C				&	\textbf{67.24} 	&	\textbf{65.71} 	&	\textbf{49.94} 	\\
			M2MRF-D				&	\underline{66.66} 	&	\underline{65.15} 	&	\underline{49.36} 	\\ \hline						
		\end{tabular}
	\end{center}
	\caption{Comparing with different RF operators on IDRiD \cite{porwal2020idrid} testing set. The first row is the baseline i.e. vanilla HRNetV2 \cite{wang2020deep}. A: (one-step/one-step), B:(one-step/cascade), C:(cascade/one-step), D: (cascade/cascade). All results are averaged over three repetitions.}
	\label{tab:RFcomparisonIDRiD}
\end{table}

\begin{table*}[tbh!]
	\begin{center}
		\begin{tabular}{|l|cccc|c|c c c c|c|c c c c|c|}
			\hline
			\multirow{2}{*}{Methods}&\multicolumn{5}{c|}{AUPR}  & \multicolumn{5}{c|}{F} & \multicolumn{5}{c|}{IoU} \\\cline{2-16}
			& EX & HE & SE & MA & mAUPR 
			& EX & HE & SE & MA &  mF 
			& EX & HE & SE & MA &  mIoU 
			\\ 
			
			\hline				
			VRT(1st)\cite{porwal2020idrid} & 71.27 & \underline{68.04} & \textbf{69.95} & \underline{49.51} & 64.69 & -& -& -& -& - &-&-&-&-&- \\
			PATech(2nd)\cite{porwal2020idrid} & \textbf{88.50} & 64.90 & - & 47.70 & - & -& -& -& -& - &-&-&-&-&- \\
			iFLYTEK(3rd)\cite{porwal2020idrid}  & \underline{87.41} & 55.88 & 65.88 & \textbf{50.17} & 64.84 & -& -& -& -& - &-&-&-&-&- \\ 
			L-seg\cite{guo2019seg} & 79.45 & 63.74 & 71.13 & 46.27 & 65.15 & -& -& -& -& - &-&-&-&-&- \\\hline   
			
			HED\cite{xie2015holistically}	&	80.81 	&	66.41 	&	68.09 	&	40.45 	&	63.94 	
			&	78.60 	&	64.33 	&	67.00 	&	38.81 	&	62.18 
			&	64.74 	&	47.43 	&	50.38 	&	24.07 	&	46.66 	
			\\
			
			DNL\cite{yin2020disentangled}	&75.12 	&	64.04 	&	64.73 	&	32.48 	&	59.09 	
			&	73.15 	&	61.87 	&	63.96 	&	32.78 	&	57.94 	
			&	57.67 	&	44.80 	&	47.03 	&	19.61 	&	42.28 	
			\\
			
			Deeplabv3+\cite{chen2018encoder}	&	81.93 	&	64.66 	&	63.04 	&	43.14 	&	63.19 	
			&	79.60 	&	61.96 	&	61.48 	&	40.57 	&	60.90 	
			&	66.10 	&	44.90 	&	44.39 	&	25.45 	&	45.21 	
			\\

			PSPNet\cite{zhao2017pyramid}	&	75.21 	&	63.36 	&	63.65 	&	32.71 	&	58.73 	
			&	73.24 	&	60.81 	&	62.83 	&	32.63 	&	57.38 	
			&	57.78 	&	43.71 	&	45.81 	&	19.50 	&	41.70 	
			\\
			
			SPNet\cite{hou2020strip}	&	64.61 	&	54.11 	&	52.14 	&	30.14 	&	50.25 	
			&	61.45 	&	44.33 	&	49.59 	&	25.66 	&	45.26 	
			&	44.40 	&	28.50 	&	33.58 	&	14.72 	&	30.30 	
			\\
			
			HRNetV2\cite{wang2020deep}	&	82.09 	&	65.50 	&	68.68 	&	43.76 	&	65.01 	
			&	\underline{79.93} 	&	62.58 	&	67.53 	&	42.49 	&	63.14 	
			&	\underline{66.57} 	&	45.56 	&	50.99 	&	26.98 	&	47.52 	
			\\
			\hline
			
			Swin-base\cite{liu2021swin}	&	81.30 	&	67.70 	&	66.46 	&	44.19 	&	64.91 	
			&	79.64 	&	\textbf{66.42} 	&	66.00 	&	44.09 	&	64.04 	
			&	66.17 	&	\textbf{49.72} 	&	49.27 	&	28.28 	&	48.36 	
			\\
			
			Twins-SVT-B \cite{chu2021Twins} &80.09 	&	63.12 	&	68.86 	&	43.27 	&	63.84 	
			&	78.56 	&	61.98 	&	\textbf{68.19} 	&	42.42 	&	62.79 	
			&	64.68 	&	44.91 	&	\textbf{51.76} 	&	26.92 	&	47.07 	
			\\
			\hline		
			
			M2MRF-A&	82.04 	&	67.80 	&	68.23 	&	47.86 	&	66.48 	
			&	79.64 	&	65.66 	&	66.68 	&	47.59 	&	64.89 
			&	66.16 	&	48.87 	&	50.03 	&	31.23 	&	49.07 
			\\
			
			M2MRF-B&	81.98 	&	67.41 	&	66.68 	&	47.91 	&	66.00 	
			&	79.57 	&	65.39 	&	65.01 	&	47.81 	&	64.45 	
			&	66.07 	&	48.58 	&	48.16 	&	31.42 	&	48.56  
			\\
			
			M2MRF-C &	82.16 	&	\textbf{68.69} 	&	\underline{69.32} 	&	{48.80} 	&	\textbf{67.24} 	
			&	79.85 	&	\textbf{66.42} 	&	\underline{67.92} 	&	\textbf{48.63} 	&	\textbf{65.71} 	
			&	66.46 	&	\textbf{49.72} 	&	\underline{51.43} 	&	\textbf{32.13} 	&	\textbf{49.94}  
			\\
			
			M2MRF-D&	82.29 	&	66.94 	&	{69.00} 	&	{48.43} 	&	\underline{66.66} 
			&	\textbf{79.97} 	&	64.88 	&	67.50 	&	\underline{48.26} 	&	\underline{65.15} 	
			&	\textbf{66.62} 	&	48.04 	 &	{50.98} 	&	\underline{31.81} 	&	\underline{49.36} 
			\\
			
			\hline
		\end{tabular}
	\end{center}
	\caption{Results on IDRiD \cite{porwal2020idrid}. M2MRF-A: (one-step/one-step), M2MRF-B:(one-step/cascade), M2MRF-C:(cascade/one-step), M2MRF-D: (cascade/cascade). The Best and Second Best results are highlighted in boldface and underlined respectively.}
	\label{tab:rst_idrid}
\end{table*}

\subsection{Results on IDRiD}
\subsubsection{Comparison to state-of-the-art RF operators} To further validate the effectiveness of our proposed M2MRF, we conduct experiments on IDRiD \cite{porwal2020idrid} and make comparisons with existing eight pairs of RF operators. Table \ref{tab:RFcomparisonIDRiD} reports the results. It shows that M2MRF-C, {i.e.}, cascade M2MRF for downsampling and one-step M2MRF for upsampling ranks first while our M2MRF-D, {i.e.}, cascade M2MRF for both downsampling and upsampling ranks second among listed competitors. Particularly, our M2MRF-C surpasses the baseline consistently in terms of mAUPR (67.24\% \textit{vs.} 65.01\%), mF (65.71\%  \textit{vs.} 63.14\%) and mIoU (49.94\% \textit{vs.} 47.52\%) by a large margin. Interestingly, two pairs of rule-based RF operators, i.e., MaxPool/Bilinear interpolation and MaxPool/Unpooling achieve better than previous learning-based RF operators. The possible reason is that it may be difficult for learning-based RF operators to preserve activations caused by tiny lesions via learning from a small local region with limited size of receptive field.

\begin{figure*}[tbh!]
	\begin{center}		
		\includegraphics[width=0.85\textwidth]{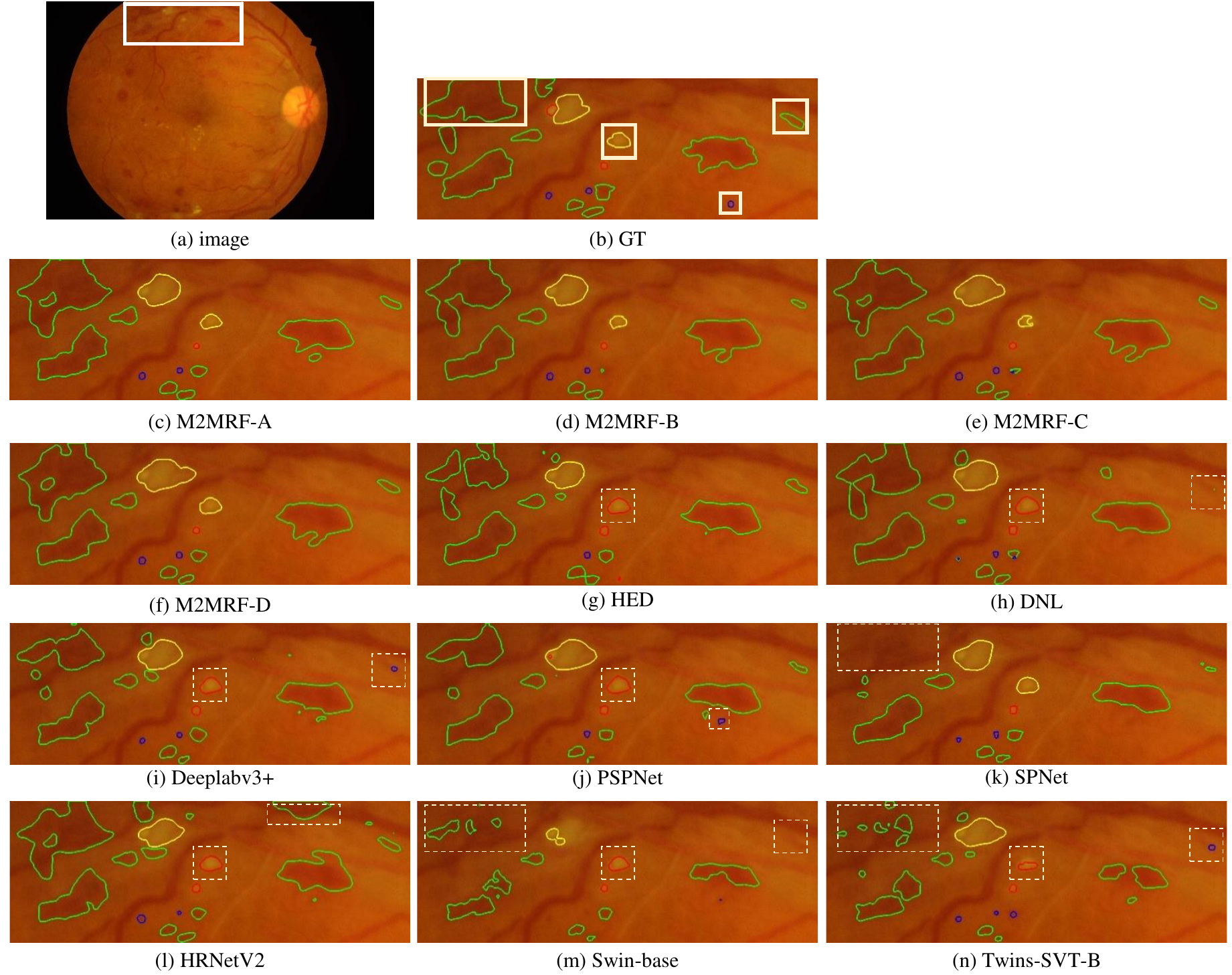}
	\end{center}
	\caption{Segmentation results on IDRiD test set \cite{porwal2020idrid}. From (a) to (n) are original image, patch with lesion annotations by experts, results by our M2MRF-A/B/C/D, HED \cite{xie2015holistically}, DNL \cite{yin2020disentangled}, Deeplabv3+ \cite{chen2018encoder}, PSPNet \cite{zhao2017pyramid}, SPNet \cite{hou2020strip}, HRNetV2 \cite{wang2020deep}, Swin-base \cite{liu2021swin} and Twins-SVT-B \cite{chu2021Twins}. Regions delineated in green, blue and red are HEs, MAs and EXs by experts or segmentation approaches. Lesions marked with solid golden boxes in (b) are challenge lesions. Four variants of HRNetV2 \cite{wang2020deep} equipped with our M2MRFs are almost able to correctly segment those challenge lesions while compared methods are prone to encounter either miss or wrong identifications which are marked with dotted golden boxes.}
	\label{fig:rstVis}
\end{figure*}

\subsubsection{Comparison to state-of-the-art segmentation methods} We compare our M2MRF with four tiny lesion segmentation methods (VRT, PATech, iFLYTEK \cite{porwal2020idrid} and L-seg \cite{guo2019seg}), six state-of-the-art CNN-based methods and two recent transformer-based methods (Swin-base \cite{liu2021swin} and Twins-SVT-B \cite{chu2021Twins}). Among four tiny lesion segmentation methods, performance of L-Seg \cite{guo2019seg} is directly borrowed from original paper and the rest are top three methods borrowed from the 2018 ISBI grand challenge. Results of six CNN-based segmentation methods and Twins-SVT-B \cite{chu2021Twins} are obtained via fine-tuning pre-trained model on ImageNet1K \cite{krizhevsky2012imagenet} with IDRiD \cite{porwal2020idrid} training set while  Swin-base \cite{liu2021swin} is ImageNet-22K. 

Table \ref{tab:rst_idrid} reports comparison results, from which we have following observations: (1) Our four variants of M2MRF surpass all the competitors consistently in terms of mAUPR, mF and mIoU; (2) They contribute significantly to MA segmentation and outperform vanilla HRNetV2 \cite{wang2020deep} by more than 4\% on AUPR, 5\% on F-score and 4\% on IoU, which indicates that our M2MRF is able to maintain more subtle activations about tiny lesions; (3) Among them, our M2MRF-C achieves best performance, which surpasses the most recent transformer-based segmentation method Swin-base \cite{liu2021swin} by 2.33\%, 1.67\% and 1.58\% in mAUPR, mF and mIoU respectively; (4) Methods specially designed for lesion segmentation achieve better in mAUPR than the compared CNN-based methods except for HRNetV2 \cite{wang2020deep}; (5) HRNetV2 \cite{wang2020deep} is a strong baseline on tiny lesion segmentation task and achieves competitive mAUPR to methods participating grand challenge as it learns high-resolution representations with detailed spatial information. We also make qualitative comparison to compared state-of-the-art methods and visualise their  results in Fig. \ref{fig:rstVis}. 

\section{Conclusion}
\label{sec:conclusion}
\color{black}
We present a simple RF operator named M2MRF for feature reassembly which unifies feature downsampling and upsampling in one framework. Our M2MRF considers contributions of long range spatial dependencies and simultaneously reassembles multiple features in a large region to multiple target features once. Thus, it is able to maintain activations caused by tiny lesions during feature reassembly. It shows significant improvements on two public tiny lesion segmentation datasets, i.e. DDR \cite{li2019diagnostic} and IDRiD \cite{porwal2020idrid}. Moreover, our M2MRF only introduces marginal extra parameters and could be used to replace arbitrary RF operators in existing CNN architectures.

\color{black}

%\vfill


\bibliographystyle{IEEEtran}
\bibliography{tmi.bib}

%\end{thebibliography}

%

\end{document}